\documentclass[
a4paper,superscriptaddress,aps,showpacs]{revtex4}
\usepackage{graphicx} 
\newcommand{\beq}{\begin{equation}}
\newcommand{\eeq}{\end{equation}}
\newcommand{\beqa}{\begin{eqnarray}}
\newcommand{\eeqa}{\end{eqnarray}}
\begin{document}
%
\title{Real clocks and the Zeno effect}
\author{I\~nigo L. Egusquiza}
\affiliation{Dept. of Theoretical Physics,
The University of the Basque Country, 644 P.B., 48080 Bilbao, Spain}
\author{Luis J. Garay}
\affiliation{Institute of Mathematics and Fundamental Physics,
CSIC, c/ Serrano 121, 28006 Madrid, Spain}
\date{January 30, 2003}
\begin{abstract}
Real clocks are not perfect. This must have an effect in our
predictions for the behaviour of a quantum system, an effect for which
we present a unified description encompassing several
previous proposals. We study the relevance of clock
errors in the Zeno effect, and find that generically no Zeno effect can
be present (in such a way that there is no contradiction with
currently available experimental data). We further observe that,
within the class of stochasticities in time addressed here, there is
no modification in emission lineshapes.
\end{abstract}
\pacs{PACS: 03.65.-w}
\maketitle 
\section{Introduction}
The problem of understanding time observables in quantum mechanics has
a long and protracted history \cite{MSE02}. One of the key
observations in the process of a better formulation of time quantities
was carried out by Misra and Sudarshan \cite{MS77}, investigating how
the measurement of lifetimes could be affected by frequent probes into
the evolution of the system under study. It should be pointed out that
Misra and Sudarshan placed their work in the context of time
observables, making explicit connections to the problem of time of
arrival, as discussed by Allcock
\cite{Allcock69,Allcock69a,Allcock69b}.

The result of Misra and Sudarshan that a continuously observed unstable
particle would never decay was associated by them to the name of Zeno
of Elea, and it is under this title that the effect or paradox is
currently known. In fact, the effect had been pointed out before in
different forms \cite{Khalfin68,Winter61}, as was indicated by Chiu,
Sudarshan, and Misra \cite{CSM77}. At any rate, the number of papers
referring to the Zeno effect or paradox increased substantially after
1977, and even more so after the crucial experiment of Itano et
al. \cite{IHBW90} (for a review of research in the topic up to 1997,
with the corresponding bibliography, see Ref. \cite{HW97}; a more
up-to-date review of bibliography can be found in Ref. \cite{FP02}).

Another intriguing aspect of time in quantum mechanics is related to
decoherence and decoherence rates, both in a general sense, and more
specifically as a source of decoherence. Milburn \cite{Milburn91}
proposed a simple modification of quantum dynamics in which the system
does not evolve continuously under unitary evolution: it undergoes a
sequence of identical unitary transformations, which take place or not
according to a Poisson distribution (i.e. the probability that there
be $n$ such transformations in a time interval $t$ is given by a
Poisson distribution). This proposal leads to decoherence, while at
the same time it conserves energy (a feature lacking in previous
models of intrinsic decoherence, such as the well known one of
Ghirardi, Rimini and Weber \cite{GRW86}).

A different aspect of stochasticity in time was put forward in
Ref. \cite{EGR98}: the fact that our clocks are not perfect implies
that incoherent superpositions of states at different instants of time
are going to be necessary to account for the state observed at a given
clock instant. The requirements for a clock to be considered good were
examined and formalized, leading to a claim of uniqueness for the
effective description of good clocks, in terms of a stationary,
Gaussian, and Markovian stochastic process.

R. Bonifacio also proposed a generalization of Liouville's equation
just by requiring that time be an stochastic variable and demanding
that the a different form of stationarity hold \cite{Bonifacio99} (see
below, eq. (\ref{convolution}) and the surrounding discussion, for the
expression of this property). He further posed a claim of uniqueness
for the probability distribution that encoded stochasticity in time,
which he asserted was the $\Gamma$ distribution.

All these three proposals shared the result that, under adequate
approximations, the systems behaving according to them would actually
follow an evolution equation for the density matrix of the form
\begin{equation}
\dot\rho(t)=-i[H,\rho(t)]-\frac{\kappa^2}{\vartheta}[H,[H,\rho(t)]]\,,\label{mastereq}
\end{equation}
where we have set $\hbar$ to 1, as we do in the following, $H$ is the
Hamiltonian of the system under consideration, described with the
density matrix $\rho$ at clock time $t$, and $\kappa$ and $\vartheta$
are constants with dimensions of time.  In the formalism of
Ref. \cite{Milburn91}, the quotient $\kappa^2/\vartheta$ is associated
to the intrinsic time step of the unitary evolutions; in
Ref. \cite{EGR98} $\kappa$ stands for the strength of the correlation
function of relative errors (the rates of increase or decrease of the
clock error at different clock times), whereas $\vartheta$ is the
correlation time for those relative errors; to be complete we mention
that the combination $\kappa^2/\vartheta$ is equivalent to the
``chronon'' of Ref. \cite{Bonifacio99}.

In fact, this kind of master equation was also known from the analysis
of heat baths coupled to the system by a term of the form $H \Gamma$,
where $H$ is the system's Hamiltonian, and $\Gamma$ some bath operator
(see for instance \cite{Carmichael93}, section 2.3). The explicit
connection between the proposal that quantum (and classical!) systems
evolve according to non-ideal clocks, on the one hand, and the heat
bath language, on the other, was shown in Ref. \cite{EGR98}.

In a recent paper \cite{Adler02}, Stephen Adler, using the language of
It\^o calculus, considered together both Zeno's effect and eq.
(\ref{mastereq}), with the result that the Zeno effect would be washed
out by the new time scale $\kappa^2/\vartheta$. Given the results of
Facchi and Pascazio (and others) \cite{FNP01,FP02}, it is rather
surprising that the Zeno effect disappears no matter the value of the
new time scale. Even more, if this equation is the result of a coupling
between the system being considered and the heat bath, the total
system (system plus bath) will of necessity present Zeno's effect.

We should mention, for the sake of completeness, that the idea of
errors in time measurements seems to be cropping up in several other
contexts by various authors (for a couple of recent examples, see
\cite{Krolikowski02,SSG02}). The motivation behind most of those
efforts seems to be either a desire to understand decoherence better,
or the search for modifications of the ordinary quantum axioms
regarding evolution. An example of this latter approach is to be found in the
series of papers \cite{Gainutdinov99,GMS02}; notice that the specific
form of non locality in time put forward in those papers leads to
radical modification of the line shape, contrary to the results of
Adler's and our own for non locality due to randomness in the
measurement of time.

The purpose of this paper is thus threefold: 1) to present a general
formalism that accounts for various different proposals in a unified
manner; 2) to examine whether Zeno's effect is indeed generically
washed away by the mere fact that time should be considered as
stochastic, or whether the result of Adler's does not indeed extend
to more general situations; 3) to resolve the different claims of
uniqueness which seem contradictory.

\section{Unified formulation}\label{sec:unified}
A good starting point is given by the initial stages of Bonifacio's
formulation \cite{Bonifacio99}, which we present here with a notation
and interpretation closer to that of Ref. \cite{EGR98}. Let
$\rho_S(s)$ represent the density matrix of system $S$ at
Schr\"odinger's ideal time $s$; that is to say, $\rho_S$ evolves
according to von Neumann's equation
\[{{{\rm d}}\over{{\rm d}s}}\rho_S(s)=-i[H,\rho_S(s)]=-i{\cal L}\rho_S(s)\,,\]
where ${\cal L}$ is the Liouvillean (super)operator.

This evolution equation is used to make predictions about the outcomes
of experiments, that we can describe in the following manner: after a
preparation stage, in which we make sure that we have set the system
in a well defined initial state, we let the system evolve a time
interval $t$ that we measure with our clock. At that instant we
measure the value of some observable of the system. Our predictions,
in general, will not concern the actual value measured in a single
instance of the experiment; they will rather provide us with
probability distributions, which will be checked by many repetitions
of the experimental procedure.

However, any clock we might use will have intrinsic uncertainties, and
the ideal time elapsed in each instance of the experiment will be
different, even if we insist on always measuring the same time
interval with our clock. We should then realize that the predictions
we are required to provide must be predictions in \textit{clock time}
($t$), not in ideal time ($s$), that is, we must have a way of
computing probability distributions for all observables at time
$t$. Through the standard arguments, this means that we can encode our
predictions in a density matrix at clock time $t$, $\rho(t)$.

The actual observed/predicted density matrix at clock time $t$ will be given by
a superposition of the density matrices at different ideal times, under the
assumption that, for different realizations of our experiment, the same
reading $t$ of our clock corresponds to different ideal time intervals,
$s$, having elapsed since the preparation of the initial state. Let us
describe this assumption by a probability density for those ideal
times $s$ at a fixed $t$, $P(t,s)$: the actual density matrix,
$\rho(t)$, will be computed as
\begin{equation}
\rho(t)=\int_0^\infty ds\,P(t,s)\rho_S(s)=
\int_0^\infty ds\,P(t,s)e^{-i s{\cal L}}\rho_S(0)\,. \label{bonifaciop}
\end{equation}
Here one simple underlying assumption is that the preparation instant
be labelled by $s=0$ and $t=0$. Another assumption is that we know for
certain that, at the instant we let the system go, the state is indeed
always the same. Note that a simple way of modifying this assumption
would be to allow $s$ to be negative. When we introduce later
$\Pi_{\mathrm{G}}(t,k)$, in eq. (\ref{pigauss}) below, we shall
actually relax this condition on $s$, as explained at that point.

By the manner in which we have justified the introduction of $P(t,s)$,
we have already demanded that it be a probability
distribution. However, an alternative way of arriving at the same
concept would be to assume eq. (\ref{bonifaciop}) and that $\rho(t)$
indeed be a density matrix (as is to be expected from the general
arguments of Gleason's theorem --- see the strengthened version
provided by P. Busch \cite{Busch99}). It would follow that $P(t,s)$
would have to be positive, and normalized to 1 when integrated over
$s$,
\[\int_0^\infty ds\,P(t,s)=1\,.\]

There is yet another important property that this probability density
should fulfill: the way in which errors could accumulate should be
independent of the instant of time at which we have started the
clock. At least it should be so for good clocks; the way new errors
are produced should be independent of the error up to that
instant. Note however that the error at one clock time $t+dt$ does
depend on the error at time $t$. If this one is very big, it is
very unlikely that at a later clock time the error could be
zero. Therefore the stationarity requirement we are now discussing
cannot be understood as stationarity for the errors in time
measurement, but rather as stationarity in the buildup of errors.

In order to obtain a mathematical statement of this stationarity
requirement, consider the following setup: a system is prepared in an
initial state $\rho_S(0)$, and is then evolved a clock time interval
$t_1$. The state at this clock time is now
\[\rho(t_1)=\int_0^\infty ds_1\,P(t_1,s_1)e^{-i s_1{\cal L}}\rho_S(0)\,.\]
By whichever means, the state is then frozen, and then we let
it evolve a further time interval $t_2$. This is another way of saying
that we repeat the evolution, but now for a different time interval,
and with an initial state that is now $\rho(t_1)$. The end result
would be the density matrix
\[
\rho(t_1+t_2)=\int_0^\infty ds_2\, P(t_2,s_2)e^{-i s_2{\cal L}}\rho(t_1)=
\int_0^\infty ds_2 P(t_2,s_2)e^{-i s_2{\cal L}}
\int_0^\infty ds_1\,P(t_1,s_1)e^{-i s_1{\cal L}}\rho_S(0)
\,.
\]
On the other hand, that freezing of the state need not be real, it is
simply a tool of our imagination, and the density matrix at the end
must be given by
\[
\rho(t_1+t_2)=\int_0^\infty ds\,P(t_1+t_2,s)e^{-i s{\cal L}}\rho_S(0)\,.
\]
This entails the following condition on the probability density $P$:
\begin{equation}
P(t_1+t_2,s)=\int_0^s ds'\,P(t_2,s-s')P(t_1,s')\,,
\label{convolution}
\end{equation}
which is the mathematical expression of the stationarity requirement.

In order to solve this functional equation, it is useful to use
Fourier transforms. Define
\[\Pi(t,k)=\int_0^\infty ds\,P(t,s) e^{i k s}\,.\]
Note for further use the following set of facts: since $P$ is a
probability distribution for the stochastic variable $s$, it follows
that $\Pi(t,0)=1$ (remember that $s$ is stochastic with respect to the
clock time which we measure); since $P$ has support only on the
positive half-line (as we are currently assuming) $\Pi(t,k)$ will have
singularities only on the lower half of the complex $k$ plane.  If all
moments in $s$ of the probability distribution $P$ existed, then
$\Pi(t,k)$ would be analytic at $k=0$.

By means of this definition, equation (\ref{convolution}) is
transformed into the algebraic equation
\[\Pi(t_1+t_2,k)=\Pi(t_1,k)\Pi(t_2,k)\,,\]
with solution
\begin{equation}
\Pi(t,k)=\left[g(k)\right]^{t/\lambda}\,,\label{solal}
\end{equation}
where the dimensionful parameter $\lambda$ has been introduced to
adimensionalize the exponent, and $g(k)$ is an arbitrary
function. Thus, $\lambda$ and $g(k)$ fully determine the
characteristic function $\Pi(t,k)$, and consequently, the probability
densities $P(t,s)$. The physical significance of $\lambda$ and $g(k)$
will become clear soon. For the time being, it is worth noticing that
$g(k)$ is the characteristic function of the probability density
$P(t,s)$ at time $t=\lambda$, and that changes in the parameter
$\lambda$ can be compensated by a change in the function $g(k)$ as
follows. Let $g_\lambda(k)$ be an arbitrary function that determines
$\Pi(t,k)$ together with the parameter $\lambda$, that is to say,
under the condition $\Pi(\lambda,k)=g_\lambda(k)$.  It is then
straightforward to see that the same distribution $\Pi(t,k)$ can be
described in terms of a new function $g_\mu(k)$ and parameter $\mu$,
such that $g_\lambda(k)=\left[g_\mu(k)\right]^{\lambda/\mu}$.

The requirement that $\Pi(t,k)$ be the Fourier transform of a probability
density, with the definition above, implies that $g(0)=1$. Similarly,
the condition on the location of the singularities also applies to
$g(k)$. Additionally, if we make the further hypothesis that all ($s$)
moments of the density $P(t,s)$ exist, this would entail that $g(k)$
would have to be analytic at $k=0$. It is also clear that $g(k)\to0$
on the real line as $|k|\to\infty$, if $P(t,s)$ is a continuous
density.

Assuming the existence of the first two moments (in $s$) of the
distribution $P(t,s)$, we can make the following interesting
statements derived from the form of $\Pi(t,k)$ expressed in
eq. (\ref{solal}). The expectation value of $s$ is given by
\[\langle s\rangle=\int_0^\infty ds\,P(t,s) s= 
-i\partial_k\Pi(t,k)\big|_{k=0}=\frac{-i g'(0)}{\lambda} \, t\,,\]
where we have taken into account that $g(0)=1$. Observe that the
expectation value is proportional to $t$, and that it is indeed
exactly $t$ if $\lambda=-ig'(0)$. It is immediate to appreciate that
the ratio $-i g'(0)/\lambda$ measures the systematic drift of the
expected values of time. The systematic drift can be eliminated by
recalibration of the clock, identified by its characteristic function,
and, indeed, the ratio $-i g'(0)/\lambda$ is invariant under the
transformations $\lambda\to\mu$ above, and pertains exclusively to the
function $\Pi(t,k)$: if it is the case that $\lambda=-ig'_\lambda(0)$,
it is then true that $\mu=-i g'_\mu(0)$.  As to the variance, one can
easily compute
\[\Delta s^2=\int_0^\infty ds\,P(t,s) 
(s-\langle s\rangle)^2=\frac{1}{\lambda}
\left(-g''(0)+g'(0)^2\right) t\,,\]
whence we see that $\Delta s\sim \sqrt{t}$ (with an adequate
dimensionful proportionality constant).

The simplest function that fits those criteria (to recap: $g(0)=1$,
analyticity at $k=0$, all singularities in the lower half plane, and
$g(k)\to0$ as $|k|\to\infty$ on the real line), and presents a
singular point, is
\[g_{\mathrm{B}}(k)={1\over{1-i k \tau}}\,,\]
where $\tau$ is a real number with dimensions of time. Notice that we
are actually choosing a family of probability densities parametrized
by $\lambda$, i.e. by the instant of time at which the characteristic
function $\Pi(t,k)$ equals $g_{\mathrm{B}}(k)$. For each member of the
family, as we shall see below, the meaning of the characteristic time
$\tau$ will be slightly different. The subscript $B$ stands for
Bonifacio: the probability density derived from this choice is
\[
P_{\mathrm{B}}(t,s)={1\over{\tau}}{{e^{-s/\tau}}\over{\Gamma(t/\lambda)}}
\left({{s}\over{\tau}}\right)^{t/\lambda-1}\,,\] which is exactly the
one put forward in Ref. \cite{Bonifacio99}. The expectation value is
$\langle s\rangle=\tau t/\lambda$, and the dispersion $\Delta
s=\tau\sqrt{t/\lambda}$. This means that among all the different
members of this family of distributions, the one labelled by
$\lambda=\tau$ is the only one for which there is no systematic drift
in the expectation value of time; in which case $\tau$ gives the rate
of growth of the dispersion of the successive probability densities
$P(t,s)$.

Another simple alternative, with a very different analytic structure
associated to periodicity, is given by
\[g_{\mathrm{M}}(k)=\exp\left(e^{i k \tau}-1\right)\,,\]
where, again, $\tau$ is a real number with dimensions of time, and the
subscript $M$ now stands for Milburn. This function
$g_{\mathrm{M}}(k)$ is an entire periodic function, bounded on the
real line (but of course not everywhere, in keeping with Liouville's
theorem). On performing the inverse Fourier transform, one is led to
the discrete probability density
\[
P_{\mathrm{M}}(t,s)=\sum_{n=0}^\infty{1\over{n!}}
\left({{t}\over{\lambda}}\right)^n
e^{-t/\lambda}\delta(s-n\tau)\,,
\] 
which is precisely that proposed in Ref. \cite{Milburn91} (eq. (2.7); in
order to make the identification between this formulation and the
original one it is convenient to use the construction of the next
section). This discrete probability density has the interpretation
provided in that paper: the probability that there be $n$ identical
unitary transformations of the density matrix, $\exp(-i\tau{\cal
L})\rho(0)$ in a time interval $t$ is given by the Poisson
distribution.  The expectation value $\langle s\rangle$ and the
dispersion are given by the same expressions as those for
$g_{\mathrm{B}}(k)$.

The two $g$ functions presented above do not of course exhaust all
possible alternatives. Just as an example, consider
\[g_{\mathrm{a}}(k)={1\over{(1-i k \tau)(1-ik\sigma)}}\,,\]
where both $\tau$ and $\sigma$ are positive time quantities (the subindex a stands for ``alternative''). Note that
if $\tau$ and $\sigma$ are equal the resulting probability would be
$P_{\mathrm{B}}(2t,s)$, which could also be phrased as
$P_{\mathrm{B}}(t,s)$ by halving the value of $\lambda$. Without loss
of generality, let $\tau>\sigma$. The resulting probability density is
\[
P_{\mathrm{a}}(t,s)=\theta(s)\frac{\sqrt{\pi}}{\Gamma(t/\lambda)}
\frac{1}{\sqrt{\sigma\tau}}
\left(\frac{s}{\tau-\sigma}\right)^{-1/2+t/\lambda}
 e^{-(\tau+\sigma) s/2\sigma\tau}
I_{t/\lambda-1/2}\left(\frac{(\tau-\sigma)s}{2\sigma\tau}\right)\,,
\]
with $I_\nu(z)$ the modified Bessel function. This probability density
serves as a counterexample to the uniqueness claim presented in
Ref. \cite{Bonifacio99}. It should be pointed out that we make no
claim whatsoever to the greater physical significance of this
distribution, as compared to $P_{\mathrm{B}}$, and they are to be
evaluated according to their fitting whatever phenomena we would like
to describe. The expectation value computes to be $\langle
s\rangle=(\sigma+\tau) t/\lambda$, while the dispersion reads $\Delta
s=\sqrt{(\tau^2+\sigma^2)t/\lambda}$.

\section{The master equation}
Direct comparison of the second integral in equation
(\ref{bonifaciop}) with the definition of $\Pi(t,k)$ tells us that we
can write the averaging process leading to the observed density matrix
at clock time $t$ in the form
\begin{equation}
\rho(t)=\Pi(t,-{\cal L})\rho_S(0)=e^{(t/\lambda)\ln\left[g(-{\cal L})\right]} 
\rho_S(0)\,.\label{exactabs}
\end{equation}
By using the eigenoperators of the Liouvillean, that is to say,
operators of the form $|n\rangle\langle m|$, where $|n\rangle$ and
$|m\rangle$ are eigenstates of the Hamiltonian, we can write the exact
evolution of the components of the density matrix in that basis: let
$\rho(t)$ be given by $\sum_{m,n}\rho_{nm}|n\rangle\langle m|$. The
components of the observed density matrix at clock time $t$ are
related to those at time $0$ by
\begin{equation}
\rho_{nm}(t)=\Pi(t,E_m-E_n)\rho_{nm}(0)\,.\label{exactcomp}
\end{equation}
This formal exact solution is very useful when analyzing simple
systems; however, it might be cumbersome in more involved situations,
and some other simplifications and limits might come in handy.  For
this reason, let us rewrite eq. (\ref{exactabs}) as a differential
equation,
\begin{equation}
\dot\rho(t)={1\over{\lambda}}\ln\left[g(-{\cal L})\right] \rho(t)\,.
\label{effective}
\end{equation}
Since $g(0)=1$ and $g(k)$ must be analytic at $k=0$, eq.
(\ref{effective}) admits an expansion whose first terms will be
\begin{equation}
\dot\rho(t)=\frac{1}{\lambda}\left(-g'(0) {\cal L}-
\frac12(-g''(0)+g'(0)^2){\cal L}^2+\ldots\right) \rho(t)\,,\nonumber
\end{equation}
or, taking into account that this expansion is actually the expansion
of the generating function of the cumulants,
\begin{equation}
\dot\rho(t)=\left(-i\frac{\langle s\rangle}{t} {\cal L}-
\frac12\frac{\Delta s^2}{t}{\cal L}^2+\ldots\right) \rho(t)\,.
\label{effectiveexp}
\end{equation}

In order for this expansion to be physically relevant it is clear that
there must be either a renormalization of the energy, or else the
first coefficient of the Taylor expansion of $g(k)$ around $k=0$ must
be given by $g'(0)=i\lambda$. In other words, the clock must be such
that it adequately tracks ideal time, with no rescaling being
necessary. Hence forward we shall assume that the statement
$g'(0)=i\lambda$ does indeed hold (which entails a relation among the
parameter $\lambda$ and the characteristic times appearing in
$g(k)$). It would be also pertinent to have some explanation for the
validity of the expansion, which both for $g_{\mathrm{B}}(k)$ and
$g_{\mathrm{M}}(k)$ reads, to second order in $k$, and making use of $\lambda=\tau$ (which is the specific form of the condition $g'(0)=i\lambda$),
\[
g_{\mathrm{B,M}}(k)= 1+ i \tau k -\tau^2 k^2+O\left(\tau^3k^3\right)
\to \dot\rho(t)=\left(-i{\cal
L}-\frac{\tau}{2}{\cal
L}^2+\ldots\right)\rho(t)\,.
\]
These examples lead us to understand the expansion
(\ref{effectiveexp}) as valid if the characteristic time of evolution
(whose inverse gives us the characteristic expansion scale of ${\cal
L}$) is much larger than the time constants that appear in the
definition of the $g$ functions. In fact, those time constants, such
as $\tau$ [or $\sigma$ and $\tau$ for $g_{\rm a}(k)$], characterize
either the period (in cases such as $g_{\mathrm{M}}(k)$) or the closest
singularities to the point $k=0$, so it is sensible to expect that
this expansion will only be valid for characteristic evolution times
larger than them, under the further demand that there is no resonant
effect.

In fact, Milburn \cite{Milburn91} provided an explicit example of the
breaking down of the expansion due to a resonant effect, by examining
the average value of an oscillator and noticing that some frequencies
would lead to the freezing of the evolution of the oscillator, namely
the harmonics of the evolution frequency $2\pi/\tau$. Indeed, one
should use in eq. (\ref{exactcomp})
\[\Pi_{\mathrm{M}}(t,E_m-E_n)=
\exp\left[\frac{t}{\tau}\left(e^{i\tau(E_m-E_n)}-1\right)\right]\,,
\]
leading to freezing of the element $\rho_{nm}(t)$ of the density
matrix if $E_m-E_n=2\pi l/\tau$, with $l$ an entire number, i.e. in the
presence of a resonance among the characteristic times of system and
real clock.

\section{Alternative stochastic realizations}
A completely different notation for the very same conceptual setup
presented in section \ref{sec:unified} is that of path integrals. This
will allow a more compact writing of some expressions, and will be of
use in establishing a different set of approximations and connections
with other models. The basic assumption is the following: a given
clock time $t$ can correspond to different ideal times $s$. These,
therefore, give rise to a stochastic function of $t$, $s(t)$, which we
will find convenient to write as $s(t)=t+\Delta(t)$; that is to say,
$\Delta(t)$ is the (stochastic) error associated with clock time
$t$. The different probability densities $P(t,s)$ lead to
corresponding probability densities for the error variable. The
stationarity requirement of eq. (\ref{convolution}) requires slightly
more delicate handling. Let us go back to the use we have made of the
probabilities $P(t,s)$. We have set an initial time $t_0$, that we
have so far denoted by $0$, at which instant we are agreed that the
elapsed ideal and clock time \textit{interval} are both zero, and the
density matrix is some initial $\rho(t_0)$ (which is for present
purposes an unevolved matrix, and therefore there is no real
distinction nor need to indicate the subscript $S$). Then some
interval $t$ has elapsed according to our clock, which reads the
instant $t_0+t$; in ideal time an interval $s$ has elapsed, associated
to an instant in ideal time $t_0+s$. The stationarity condition means
that we should not care whether $t_0$ is 5 or 55, in order to
determine the probability that an interval $s$ in ideal time has
elapsed when the clock interval is $t$.  This condition can thus be
translated into the following one on the error stochastic variable:
that the \textit{relative} error be a stationary stochastic
process. In other words, that $\Delta(t)$ be written as $\int_0^t
d\xi\,\alpha(\xi)$, and that this $\alpha(t)$ be
stationary. Furthermore, since no negative time interval can ever
elapse, we need to have $\alpha(t)\geq-1$ always (this was also
explained in \cite{EGR98} as due to ``good'' causality: if it were
possible to have relative errors smaller than $-1$, we would see the
clock going backwards) . We see that very often it is more convenient
to deal with the relative error process, and we shall write our path
integrals in terms of this stochastic function, with weight ${\cal
P}[\alpha]$. In this manner, we recover the probability density
$P(t,s)$ as a sum over all possible sequences of relative errors (this
is what $\int {\cal D}\alpha$ will stand for), weighted with
probability functional ${\cal P}[\alpha]$, restricted to those
sequences for which it happens that $t+\Delta(t)$ equals $s$:
\[P(t,s)=\int{\cal D}\alpha\,{\cal P}[\alpha]\,\delta(t+\Delta(t)-s)\,,
\qquad\Delta(t)=\int_0^t
d\xi\,\alpha(\xi)\,. \]

Since $\alpha(t)\geq-1$, the simplest possible choice for the
functional weight ${\cal P}[\alpha]$, namely a Gaussian weight, is not
feasible. Another way to see that this is indeed the case is to
compute the Fourier transform of the probability density
\[\Pi(t,k)=\int{\cal D}\alpha\,{\cal P}[\alpha]\,e^{i kt}e^{ik\Delta(t)}\]
for a Gaussian weight, ${\cal
P}_{\mathrm{G}}[\alpha]=\exp\left[-\frac12\int d\xi d\xi'
\alpha(\xi)c^{-1}(\xi-\xi')\alpha(\xi')\right]$ 
(for simplicity we have taken $\langle\alpha(t)\rangle=0$; this
actually implies that the average value of $s$ is $t$, thus choosing
the correct scaling):
\begin{equation}
\Pi_{\mathrm{G}}(t,k)=
\int{\cal D}\alpha\,
e^{-\frac12\int d\xi d\xi' \alpha(\xi)c^{-1}(\xi-\xi')\alpha(\xi')}\,
e^{i kt}e^{ik\Delta(t)}=
e^{i kt}e^{-\frac{k^2}2\int_0^t d\xi\int_0^t d\xi' c(\xi-\xi')}=
e^{ikt} e^{-k^2f(t)}\,,\label{pigauss}
\end{equation}
where $c(\xi)$ is the correlation function of relative errors,
\[\langle\alpha(t)\alpha(t')\rangle=\int{\cal D}\alpha\,{\cal P}_{\mathrm{G}}[\alpha]\alpha(t)\alpha(t')=c(t-t')\,,\]
$c^{-1}(\xi)$ its functional inverse, and $f(t)$ is explicitly defined
in the last equality of eq. (\ref{pigauss}).

Condition (\ref{solal}) can only hold for a Gaussian functional
weight, as can be seen from comparison with eq. (\ref{pigauss}), if
$g(k)=\exp(i k\lambda)$, which means that the clock is perfect and
always gives perfect ideal time, i.e. $P(t,s)=\delta(t-s)$.

On the other hand, a real clock could perfectly well go back in time,
in which case the restriction above would no longer hold. The
stationarity requirement (\ref{convolution}), would need not hold
either, since the imaginary process by which we arrived at it could
now entail retrodictions. The generic form of the Fourier transform of
the probability density for a Gaussian functional weight centered at
$\langle\alpha(t)\rangle=0$ (as we must demand to avoid systematic
slipping of the clock) is of the form
\[
\Pi_{\mathrm{G}}(t,k)=e^{ikt} e^{-k^2f(t)}\,,
\]
as computed above,
and the corresponding probability density
\[P_{\mathrm{G}}(t,s)=\frac{1}{2\sqrt{\pi f(t)}}e^{-(t-s)^2/4 f(t)}\,,\]
where $s$ is no longer restricted to being positive. The Gaussian
process leads to an effective evolution equation
\[\dot\rho_{\mathrm{G}}(t)=\left(-i{\cal L}-
\dot f(t) {\cal L}^2\right)\rho_{\mathrm{G}}(t)\,, \]
which is precisely of the form of the master equation
(\ref{mastereq}), with a time dependent coefficient. The reason for
the coefficient of the linear term in ${\cal L}$ to be $-i$ is that
the average value $\langle\alpha(t)\rangle$ is zero. If it were not
zero, then this coefficient would acquire a factor
$1+\langle\alpha(t)\rangle$. Concentrating now on the quadratic term
in ${\cal L}$, let us point out that the case of a constant
coefficient corresponds to the Wiener process for the absolute
errors. Notice that all the higher order cumulants vanish and the
master equation is exactly given by the first two terms.

This signals that whenever we can simulate the evolution of the
quantum system by the master equation (\ref{mastereq}), we can perform
multi-time (and single time) computations using a Gaussian Markovian
stationary functional weight ${\cal P}[\alpha]$: the
Ornstein-Uhlenbeck process, in fact, in the stationary regime. Let us
be more specific about this issue. If we indeed were to use the
Ornstein-Uhlenbeck process for the relative errors, with correlation
function $\langle\alpha(t)\alpha(t')\rangle=
(\kappa/\vartheta)^2\exp(-|t-t'|/\vartheta)$, the function $f(t)$
would be
\[f_{\mathrm{OU}}(t)=
\kappa^2\left(\frac{t}{\vartheta}-1+e^{-t/\vartheta}\right)\,,\]
and the effective evolution equation would read
\[\dot\rho_{\mathrm{OU}}(t)=
\left(-i{\cal L}-\frac{\kappa^2}{\vartheta}\left(1-e^{-t/\vartheta}\right) 
{\cal L}^2\right)\rho_{\mathrm{OU}}(t)\,, \] which indicates that
there is a transient effect up to times of the order of the
correlation time for the process, $\vartheta$. Notice that at $t=0$
there are no non-Liouvillean terms, due to the specific character of
the transient of the correlation function. As we shall see, this will
be particularly relevant in Zeno's effect. After times of order $\vartheta$, the evolution is dictated by the master equation (\ref{mastereq}). For the sake of later convenience, observe that the characteristic function of the master equation, under the identification $\tau=2\kappa^2/\vartheta$, is given by
\[\Pi_{\mathrm{m.e.}}(t,k)=e^{ik t - k^2\tau t/2}\,.\]
Quite obviously, the subscript m.e. stands for ``master equation.''
\section{The Zeno effect}
The generality of Zeno's effect has been shown in very many forms. An
early clear description of the general character of the effect was
provided by Chiu, Sudarshan, and Misra \cite{CSM77}: if the
Hamiltonian is bounded from below, and the initial state is such that
the expectation value of the Hamiltonian is finite, then the
derivative of the survival probability with respect to time at the
initial instant is zero, which entails that the decay of the survival
probability is slower than any exponential. The survival probability
at ideal time $s$ is given by $\mathrm{Tr}\left\{\rho(0)\left[\exp(-i
s {\cal L})\rho(0)\right]\right\}$. As, under these conditions, this
survival probability has no linear term, close to $s=0$, on performing
sufficiently frequent measurements we find that the survival
probability at any later time is equal to 1: the system is confined to
its initial state. On the other hand, if the survival probability did
indeed have a linear term in $s$, then the evolution under frequent
measurements would be an exponential decay, with the decay constant
given by the coefficient of the linear term.

We shall now investigate whether the uncertainties in time do
eliminate Zeno's effect from taking place. The quantity we want to
investigate is the survival probability, that is, given an initial
state $\rho(0)$, we should compute
\[p(t)=\mathrm{Tr}\left[\rho(0)\rho(t)\right]\,,\]
where we measure the time elapsed with a real clock, and, therefore
[see eq. (\ref{exactabs})]
\[p(t)=
\mathrm{Tr}\left\{\rho(0)\left[\Pi(t,-{\cal L})\rho(0)\right]\right\}\,.\]
Note that the objects which include the Liouvillean are
super-operators, not operators. This entails a slight complication of
notation, which we will fix by requiring that super-operators act on
everything on their right.

For measurements of time intervals that satisfy the stationarity
constraint (\ref{convolution}), the linear term in clock time of the
survival probability is given by
\[\frac{1}{\lambda}\mathrm{Tr}\left\{\rho(0)
\ln\left[g(-{\cal L})\right]\rho(0)\right\}\,,\]
which, in terms of the energy basis, can be written as
\[\frac{2}{\lambda}\sum_{n>m}\left|\rho_{nm}(0)\right|^2 
\ln\left|g(E_n-E_m)\right|\,.\]
Unless this quantity is zero, the quantum Zeno effect will not take
place (notice by the way in the previous computation that, by
construction, $g(k)^*=g(-k)$ for real $k$.).

Let us assume the validity of the expansion in
eq. (\ref{effectiveexp}), with $\langle s\rangle=t$, and further that
the initial state is a pure state, $\rho(0)=|a\rangle\langle a|$. In
such a situation, the linear term will have a leading term of the form
\[-\frac{\Delta s^2}{2t}\mathrm{Tr}\left[\rho(0)
{\cal L}^2\rho(0)\right]=
-\frac{\Delta s^2}{t}\left(\Delta H\right)^2\,, \] where $\left(\Delta
H\right)^2=\langle a|H^2|a\rangle-\langle a|H|a\rangle^2$. That is to
say, the small time survival probability will be
\[p(t)= 1-\frac{\Delta s^2}{\tau_Z^2}+\ldots\,,\]
with $\tau_Z=1/\Delta H$.  No Zeno effect survives: frequent
measurements of a system will not maintain it in the initial state,
since $\Delta s^2\sim t$. There will be an exponential decay no matter
how fast the measurements are! Admittedly with a very small decay
constant, but exponential nonetheless.  We can reobtain this result in
a slightly different fashion, by noting that all terms of the Taylor
expansion of the (ideal) survival probability in ideal time $s$
contribute to a linear term in $t$ when we perform the averaging with
$P(t,s)$: all cumulants are proportional to $t$, as we have seen, and
this is itself a consequence of the stationarity property.

For general clocks, that do not necessarily fulfill stationarity
condition (\ref{convolution}), the term linear in time will be of the
form $\mathrm{Tr}\left[\rho(0)\partial_t\Pi(0,-{\cal
L})\rho(0)\right]$, generically non vanishing. In the Gaussian case, which is not stationary in the sense of eq. (\ref{convolution}), with initial
pure state, leads to
\[p_{\mathrm{G}}(t)=1-\frac{2\dot f(0)}{\tau_Z^2}t+O(t^2)\,.\]
Thus a very special situation is associated with the Ornstein-Uhlenbeck
clock, for which $\dot f(0)=0$, and Zeno's effect survives. Note
however the exceptional character of this case, due to the specific form of the transient. 

We have thus shown that clock errors generically wash out any possible
Zeno effect. This being the case, how can we explain the experimental
results of Itano et al. \cite{IHBW90}? Forgoing an analysis in terms
of the full three-level system coupled to the electromagnetic field
\cite{BM95,Ballentine91,BB91,FS91,BH96a}, and neglecting optical pumping due
to the measuring laser, we can concentrate on the following conceptual
setup: consider a two level system undergoing a $\pi$-pulse Rabi
oscillation. At regular intervals the system is queried as to the
state it is in, whether the first or the second level. Formally, at
regular intervals the coherences of the density matrix, $\rho_{12}$
and $\rho_{21}$ are set to zero. In the standard analysis, the
evolution between measurements is ordinary unitary Schr\"odinger
evolution with the Hamiltonian
$H=\frac\Omega2\pmatrix{0&1\cr1&0}$. Given an initial density matrix
$\rho(0)=\pmatrix{b&0\cr0&1-b}$, and measurements at intervals
$\pi/(n\Omega)$, the probability of finding the system in the second
level $\pmatrix{0\cr1}$ after time $\pi/\Omega$ is
\[p_2\left(\frac\pi\Omega\right)=\frac12+\left(\frac12 - b\right)
\cos^n\left(\frac\pi{n}\right)\,.\]
If the evolution and measurement process were to take place according
to a clock with characteristic function $\Pi(t,k)$, the population of
the second level would read
\[p_2\left(\frac\pi\Omega\right)=\frac12+\left(\frac12 - b\right)
\mathcal{C}^n\left(\frac\pi{n}\right)\,,\]
where
\[\mathcal{C}(t)=\frac12\left[\Pi(t,\Omega)+\Pi(t,-\Omega)\right]\,.\]
It is worth mentioning that in the case of a perfect clock, for which
$\Pi_\mathrm{perfect}(t,k)=\exp(ik t)$ both expressions coincide, as
they should.

In order to illustrate this result numerically, let us consider the
characteristic function of the master equation, which leads to 
\[p_2\left(\frac\pi\Omega\right)=\frac12+\left(\frac12 - b\right)
e^{-\Omega\tau\pi/2}\cos^n\left(\frac\pi{n}\right)\,.\] For the system
examined by Itano et al., $\Omega=\pi/256\ \mathrm{ms}^{-1}$, and
$\tau$ should be larger than $10^{-5}$ s for it to have any noticeable
effect. As a matter of fact, $\tau$ should be larger than $10^{-4}$ s
for the effect to compete with optical pumping due to the measuring
laser. On the other hand, the coherence times mentioned in
\cite{IHBW90} (550 s) lead to $\tau\le10^{-4}$ s; additionally, the
precisions mentioned for time quantities in their experiment are of
the order $10^{-5}$ s or better. It follows that the inhibition of
transitions from level 1 to level 2 due to frequent measurement is
still in place, even though no ``perfect'' Zeno effect could take
place if the clocks themselves are not perfect.

\section{Examples:  oscillating and decaying systems}

Consider an oscillating system in initial state $|a\rangle$ evolving under the
Hamiltonian $H$,
\[
H=\pmatrix{0&\Omega&0\cr\Omega&0&K\cr0&K&0\cr}\,,\qquad|a\rangle=\pmatrix{1\cr0\cr0}\,.
\]
It is easy to
compute the exact survival probability under unitary evolution with
the preceding Hamiltonian, and then perform the averaging over clock
errors, leading to
\[
p(t)=\frac1{\Omega'^4}\left\{K^4+\frac{\Omega^4}2 + 
2 K^2\Omega^2\mathrm{Re}\left[\Pi(t,\Omega')\right]+
\frac{\Omega^4}{2}\mathrm{Re}\left[\Pi(t,2\Omega')\right]\right\}\,,
\]
where $\Omega'=\sqrt{\Omega^2+K^2}$.

The small time expansion of this exact evolution in the case of the
decohering master equation (\ref{mastereq}), with
$\tau=2\kappa^2/\vartheta$, is quite simply
\[p_{\mathrm{m.e.}}(t)\sim1-\Omega^2\tau t+ O(t^2)\,.\]
It is immediate to see that there will be no full Zeno effect, since
there is a linear term in the expansion.

If we were to use $g_{\mathrm{M}}(k)$, the full result would be
\[
p_{\mathrm{M}}(t)=\frac1{\Omega'^4}\left[K^4+\frac{\Omega^4}2 + 
2 K^2\Omega^2 e^{-(1-\cos\Omega'\tau)t/\tau} 
\cos\left(\frac{t}{\tau}\sin\Omega'\tau\right)+ 
\frac{\Omega^4}{2}e^{-(1-\cos2\Omega'\tau)t/\tau} 
\cos\left(\frac{t}{\tau}\sin2\Omega'\tau\right)\right]\,.
\]
A particularly interesting aspect of this expression is that whenever
$\Omega'=2n\pi/\tau$ the Zeno effect reappears: the motion of the
system is indeed frozen, as Milburn pointed out for different reasons.

Finally, for good Ornstein-Uhlenbeck clocks we find
\begin{equation}
p_{\mathrm{OU}}(t)=
\frac1{\Omega'^4}\left(K^4+\frac12 \Omega^4 
+ 2 K^2\Omega^2 e^{-\kappa^2\Omega'^2
\left(t/\vartheta-1+e^{-t/\vartheta}\right)}
\cos(\Omega' t) 
+ \frac12 \Omega^4 e^{-4\kappa^2\Omega'^2
\left(t/\vartheta-1+e^{-t/\vartheta}\right)}
\cos(2\Omega' t)\right)
\,.
\end{equation}
As soon as $t$ is substantially bigger than the correlation time of
the OU clock the two survival probabilities computed with the master
equation and with the OU probability weight coincide. However, for
small times their behaviour is radically different: there is no linear
term in $t$ in the expansion of $p_{\mathrm{OU}}(t)$, in keeping with
the general formal result presented above.

Let us now perform an
analogous computation for a model of a decaying system, in such a way
that we can extrapolate without difficulty to more general decaying
processes. On the by-side we will obtain an expression for the line
shape, confirming generically the result of Adler's that it is not
modified by phase decoherence \cite{Adler02}.

Consider thus a system with a discrete orthogonal basis
$\{|a\rangle\}\cup\{|\omega\rangle\}_\omega$, where $\omega$ takes
values in some discrete set of frequencies, and such that the
Hamiltonian can be written as $H=H_0+V$, with
\begin{equation}
H_0=\omega_a|a\rangle\langle a|
+\sum_\omega\omega|\omega\rangle\langle\omega|
\,,\label{hamildec}
\end{equation}
and the only nonzero elements of $V$ being $\langle a|V|\omega\rangle$
and their complex conjugates.  The initial state will be the pure
state $|a\rangle$. This system is a simplified model of decay from
this pure state to the rest of Hilbert space.

By using the resolvent, $G(z)=(z-H)^{-1}$, the exact Dyson-Schwinger's equation
\[G(z)=G_0(z)+G_0(z)VG(z)\]
for this system  can be solved to
\begin{eqnarray}
G_a(E)& =&  
\frac{1}{E-\omega_a-\Sigma_a(E)}\,,\label{sigmaa}\\
G_\omega(E) &=& \frac{\langle\omega|V|a\rangle}{(E-\omega)(E-\omega_a-\Sigma_a(E))}\,,\label{sigmag}
\end{eqnarray}
where
\[G_a(z)=\langle a|G(z)|a\rangle\,,\quad 
G_\omega(z)=\langle\omega|G(z)|a\rangle\,,\]
and
\begin{equation}
\Sigma_a(E)=
\sum_\omega\frac{|\langle a|V|\omega\rangle|^2}{E-\omega}\label{selfener}
\end{equation}
is the exact self-energy for this model.  In more general models, it
is still the case that we can write the expectation value of the
resolvent in state $|a\rangle$ in terms of the self-energy
$\Sigma_a(E)$, as expressed in eq. (\ref{sigmaa}). The change will
come about because of the modifications of the self-energy, which will
no longer be determined by eq. (\ref{selfener}). The transition
quantity $G_\omega(E)$ will also have a different expression.

The small time behaviour of the survival amplitude in ideal time $s$
is determined by the large energy behaviour of $G_a(E)$, and,
consequently, of the large energy behaviour of the self-energy. In our
specific example it is very easy to see that, for energies large in
comparison to all $\omega$ values,
\[\Sigma_a(E)\approx \frac1{E}\sum_\omega|\langle a|V|\omega\rangle|^2= 
\frac1{E}\left(\langle a|H^2|a\rangle-\langle a|H|a\rangle^2\right)=
\frac{1}{\tau_Z^2 E}\,.
\]
Under general assumptions, and for large energies, the self-energy
function has exactly the same structure $\Sigma_a(E)\sim 1/\tau_Z^2 E$
for generic models. Then, since the survival probability amplitude ${\cal
A}(s)$ is determined by the resolvent and hence by the self-energy
through
\[{\cal A}(s)= \frac{i}{2\pi}\int_B dE\,e^{-i E
s}G_a(E)\,,\] where by $B$ here we denote the adequate integration
path, the survival probability at small ideal time $s$ is
\cite{FP02}
\beq 
p_{\mathrm{ideal}}(s)=\left|{\cal A}(s)\right|^2\sim1+\frac{2}{4+\omega_a^2\tau_Z^2}
\left[\cos\left(\sqrt{4+\omega_a^2\tau_Z^2}\,
\frac{s}{\tau_Z}\right)-1\right]\,.\label{shorttprob}
\eeq
On performing the averages for Eq. (\ref{shorttprob}), we obtain
\[
p(t)\sim1+\frac{2}{4+\omega_a^2\tau_Z^2}
\left[\mathrm{Re}\Pi(t,\sqrt{\omega_a^2+4/\tau_Z^2})-1\right]\,.
\]
The objection might be posed that small $t$ does not imply that $s$ is
small (the comparison term in order to state the smallness or
otherwise of these dimensionful quantities is always taken to be the
characteristic evolution time of the system).  However, we have seen
above that clocks that satisfy the stationarity requirement of
eq. (\ref{convolution}) display an average value of $s$ that tracks
$t$, and that the dispersion is given as $\Delta s^2\sim t$. It then
follows that indeed we can approximate small $t$ behaviour by
extracting the small $s$ behaviour and then averaging.

We see again that through this method we recover again that there is a
linear term, and that the disappearance of Zeno's effect is completely
generic.  An exception can be found in the OU clock case, for which
\[
p_{\mathrm{OU}}(t)\sim 1-\frac{t^2}{\tau_Z^2}\,,
\]
as we already know.

We shall now study the \textit{large} time behaviour of the
system. Assume now that the states orthogonal to the initial one,
$|a\rangle$, form a continuum, in the manner postulated by Weisskopf
and Wigner
\cite{WW30}, so that an imaginary part can arise for the poles of
$G_a(E)$. Then, for large times, the dominant behaviour, both for the
survival amplitude and for the probability amplitude to find the
system in a state different from the initial one, will be determined
by the pole of $G_a(E)$ closest to the real axis. Assume there is only
one such relevant pole, of the form $\omega_p-i\gamma/2$.  The
contribution of this simple pole to the survival amplitude is
\[
{\cal A}_{\mathrm{p}}(s)=
\sqrt{{\cal Z}_{\mathrm{p}}} e^{-i \omega_p s-\gamma s/2}\,,
\]
where ${\cal Z}_{\mathrm{p}}$ is the relevant residue. Hence the
contribution of the simple pole to the survival probability at ideal
time $s$ becomes
\[
p_{\mathrm{p, ideal}}(s) ={\cal Z}_{\mathrm{p}} e^{-\gamma s} \,.\]
If this were the only contribution to the survival probability, it
would be immediate to conclude that at clock time $t$ the survival
probability would read
\[p_{\mathrm{p}}(t)={\cal Z}_{\mathrm{p}}\Pi(t,i \gamma)\,.\]
For large times it is to be expected that the main contribution of
real clocks will be given by the master equation approximation. It
results that
\[p_{\mathrm{p}}(t)\approx{\cal Z}_{\mathrm{p}} 
e^{-\gamma(1-\gamma\tau/2)t}\,,\] 
which means that the long times delay is slowed down as an effect of
clock errors, the new decay constant being given by
$\gamma(1-\gamma\tau/2)$. As the half-life and the error dispersion
parameter approach, the long time decay will become slower!

Under the same approximation (single simple pole or
Weisskopf--Wigner's approximation), we have for the probability
amplitude of states orthogonal to the initial one, at very late times,
and for the specific Hamiltonian (\ref{hamildec}),
\[  
{\cal B}_\omega(s)=\langle\omega e^{-i s H}|a\rangle
\sim\frac{\langle\omega|V|a\rangle\left(e^{-i\omega_p s} e^{-\gamma s/2} 
- e^{-i\omega s}\right)}{\omega_p-\omega-i\gamma/2}\,,
\]
whence it follows that the probability of finding state
$|\omega\rangle$ at (sufficiently large) clock time $t$ is
\begin{eqnarray}
p_\omega(t)&=&\left|{\cal B}_\omega(s)\right|^2\nonumber \\
&=&
\theta(s)\frac{\left|\langle\omega|V|a\rangle\right|^2}
{\left(\omega_p-\omega\right)^2+\gamma^2/4} 
\left(1+\Pi(t,i\gamma)-\Pi(t,\omega-\omega_p+i\gamma/2)-
\Pi(t,\omega_p-\omega+i\gamma/2)\right)\nonumber\\
&\to&
\frac{\left|\langle\omega|V|a\rangle\right|^2}
{\left(\omega_p-\omega\right)^2+\gamma^2/4}\,,\nonumber
\end{eqnarray}
which means that the line shape, within the approximation carried out
(Weisskopf-Winger approximation, see ref. \cite{WW30}) does not change
at all because of the errors in the clocks. We have not actually fully
proved this statement: we have not justified that the use of the
long-time approximation in ideal time is enough before taking the
average. However, for probability densities that lead to condition
(\ref{convolution}), the average value of $s$ is proportional to $t$,
and its quadratic dispersion goes instead with $\sqrt{t}$. It follows
that one can carry over the approximation of large ideal times to
large Schr\"odinger time.

\section{Conclusions}
We have shown using a general description of errors in clocks, 1) that
decoherence, although a slow process if the clocks are good, starts
taking place immediately for generic models of clocks; 2) that, as a
consequence, Zeno's effect would never fully freeze the system in its
initial state, for generic clocks. 3) We have found two kinds of
exception to this result: whenever the decoherence itself freezes the
state, as happens for some frequencies in the case put forward by
Milburn, and in the Ornstein-Uhlenbeck clock. 4) That, under the
assumptions of real clocks put forward in \cite{EGR98} and in this
paper, there is no change in the line shape due to non locality in time.

\begin{acknowledgments}
This work has been supported by Ministerio de Ciencia y Tecnolog\'\i a
(AEN99-0135, BFM2000-0816-C03-03, BFM2001-0213, BFM2002-04031-C02-02,
FPA2002-02037), UPV-EHU (00039.310-13507/2001), and the Basque
Government (PI-1999-28).  I.L.E. is grateful to M.A. Valle,
J.L. Ma\~nes and J.G. Muga for useful comments.
\end{acknowledgments}

\end{document}